\documentstyle[pre,aps,preprint]{revtex}

\begin{document}
\draft
\tighten

\title{Fluid Induced Particle Size Segregation in Sheared Granular
  Assemblies}

\author{Sitangshu Bikas Santra$^1$, Stefan Schwarzer,$^{1,2}$\cite{email}\\
 and Hans Herrmann$^{1,2}$}
\address{$^1$Laboratoire de Physique M\'ecanique 
  des Milieux H\'et\'erog\`enes,\\ Ecole Sup\'erieure de Physique et
  Chimie Industrielles, 75231 Paris, Cedex 05, France}
\address{$^2$ Institut f\"ur Computeranwendungen, Universit\"at Stuttgart,\\
 Pfaffenwaldring 27, 70569 Stuttgart, Germany}

\date{\today}
\maketitle

\begin{abstract}
  
  We perform a two-dimensional molecular-dynamics study of a model for
  sheared bidisperse granular systems under conditions of simple shear
  and Poiseuille flow.  We propose a mechanism for particle-size
  segregation based on the observation that segregation occurs if the
  viscous length scale introduced by a liquid in the system is smaller
  than of the order of the particle size. We show that the ratio of
  shear rate to viscosity must be small if one wants to find size
  segregation. In this case the particles in the system arrange
  themselves in bands of big and small particles oriented along the
  direction of the flow. Similarly, in Poiseuille flow we find the
  formation of particle bands. Here, in addition, the variety of time
  scales in the flow leads to an aggregation of particles in the zones
  of low shear rate and can suppress size segregation in these
  regions. The results have been verified against simulations using a
  full Navier-Stokes description for the liquid.

\end{abstract}  


\section{Introduction}

If a mixture of particles and liquid is sheared, either in a
continuous fashion or by periodic excitation of the system, a host of
structural rearrangements in the mixture are known to occur. Most of
the systems studied experimentally involve mono-disperse
suspensions~\cite{Noetinger86,Ackerson88,t:Gondret94}. For
concentrated suspensions and oscillatory shearing the formation of
layers in the iso-velocity planes and an additional arrangement in
lines of particles perpendicular to the velocity
field~\cite{t:Gondret94} has been observed.  In addition, simulations
of Taylor-Couette flow in the viscous regime confirm the organization
of the particles in the suspension in layers oriented along the
iso-velocity planes~\cite{Bossis84}. Numerically, cluster formation
has been found in these simulations within the formed planes. Although
for bidisperse suspensions similar organization patterns have been
found~\cite{t:Gondret94} and size-distribution induced instabilities
have been observed in sedimentation experiments~\cite{Whitmore55},
simulations here so far have not found clear indications of size
segregation under shear.

%Particle size segregation is an important phenomenon that is known to
%occur under several conditions in dry granular flows \cite{}. 
%For example, in a vibrated mixture of small and big particles, the big
%particles tend to ascend to the top surface of the mixture. Whereas the 
%phenomenon as such is well verified in experimental and numerical 
%studies, the mechanism of the size separation is still disputed;
%possible explanations range from entropic effects that favour the
%aggregation of small particles underneath big ones or convectional driving
%mechanisms. Moreover, the situation is complicated because sometimes
%it is not clear to which extent the presence of a fluid phase, e.g. 
%air, modifies the collective behavior of the grains. For example, the
%phenomenon of heaping in vibrated granular media has been attributed
%to the presence of air. 

In this paper, we want to investigate 
particle size segregation of a granular medium under shear flow
conditions in two dimensions in the presence of a liquid. The
mechanism for size segregation in this model is basically of geometric
nature and may generalize to three dimensions.  We concentrate on pipe
and laminar shear flows described by local shear rates $\dot{\gamma}
\equiv (\partial/\partial y) v_x$, where we have in mind (see Fig.
\ref{f:sketch}) that the flow is translationally invariant in $x$
direction and that its speed varies in $y$ direction.

\section{The model}
 
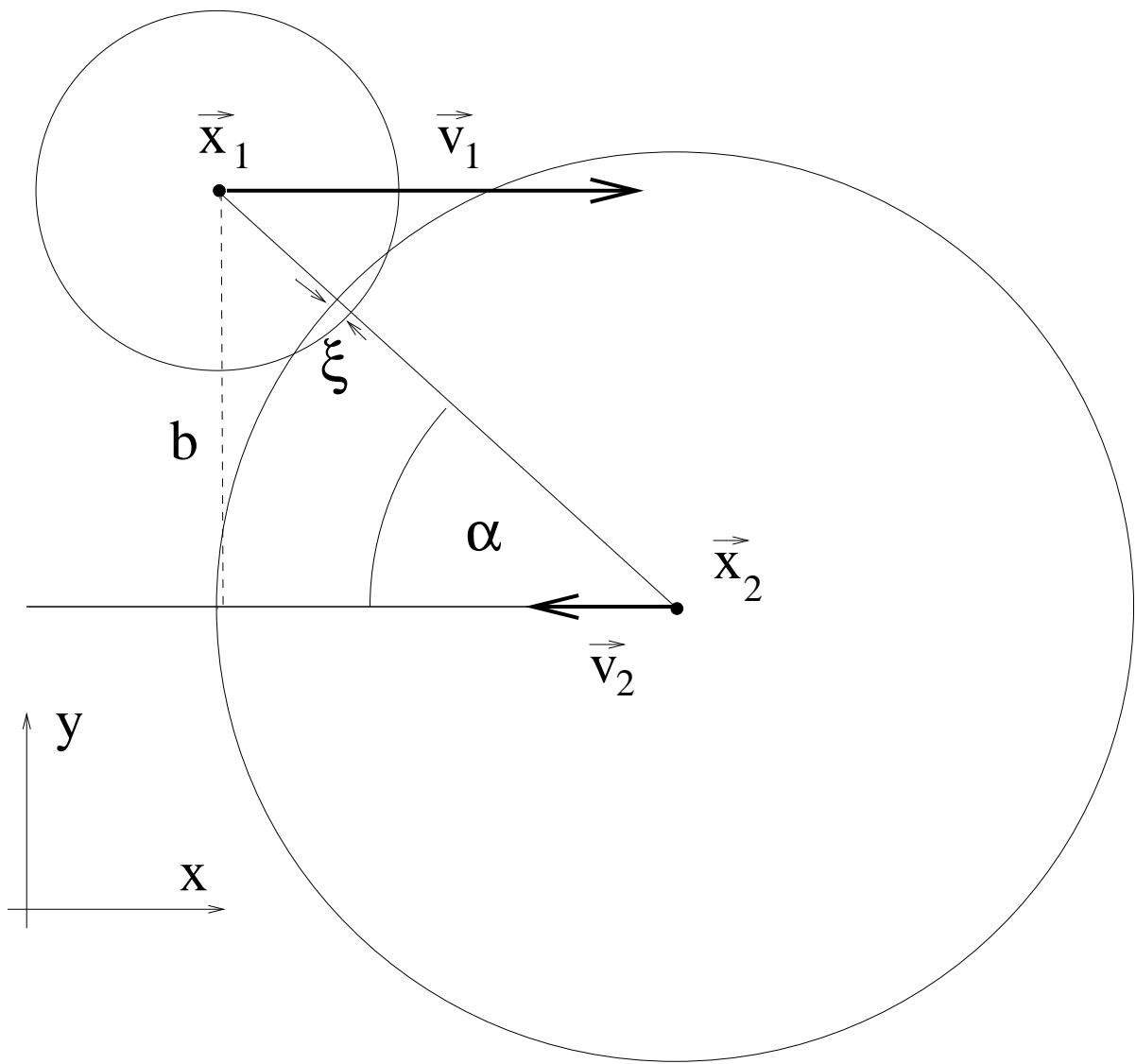
\begin{figure}
%\centerline{\includegraphics[width=.9\textwidth,angle=-90]{impact.eps}}
%\bigskip
\caption{\label{f:impact}
  Geometry of the collision between two disks in the center of mass
  frame. The particles are assumed to initially move parallel to the
  flow, such that laboratory from and center of mass frame are
  aligned.
  }
\end{figure}

For the sake of simplicity we assume the particle size distribution to
be bidisperse, i.e., our system consists of a number $n_s$ of small
and a number $n_b$ of big disk shaped particles with radii $r_s$ and
$r_b,$ respectively. These particles are initially distributed
randomly over a two dimensional plane with a specified area fraction
$c$. The overall area fraction $c$ is the sum of the area fractions of
big and small particles, $c = c_s + c_b$.  Under viscous conditions
and when two particles are nearly in contact, lubrication effects will
introduce a strong velocity dependent damping that inhibits particle
collisions.  In our model, we allow for small particle overlaps, but
between overlapping particles strong repulsive forces and and a
velocity proportional damping are introduced. I.e, in normal direction
acts a damped harmonic force $f_n,$
\begin{equation} \label{e:interaction}
f_n = - \mu \xi + 2 \mu \kappa v_n, 
\end{equation}
where $\mu$ is the reduced mass of the pair, $\xi \equiv |{\bf x}_1 -
{\bf x}_2| - (r_1 + r_2)$ is the virtual overlap of the two particles
located at ${\bf x}_1$ and ${\bf x}_2$ (Fig. \ref{f:impact}), $v_n$ is
the normal component of the relative velocity, and $\kappa$
parameterizes a velocity proportional damping. The form of Eq.
(\ref{e:interaction}) is such that the {\it restitution coefficient}
$e$, the ratio of the normal velocities after and before the impact,
is the same for all binary collisions, independent of mass and
velocity. The restitution $e$ is related to $\kappa$ by $\kappa \equiv
1/ \sqrt{1 + (\pi/\log(e))^2}$ and is chosen to be 0.8 in this work.
Equation (\ref{e:interaction}) is in dimensionless form, the chosen
length scale is the average radius $\bar{r} \equiv (n_b r_b + n_s r_s)
/(n_b + n_s),$ the mass unit is the mass of a particle of average
radius, and the time unit is such that the ``spring constant'' of a
pair contact becomes unity and thus does not appear in
Eq.~(\ref{e:interaction}). More precisely, in these units an elastic
contact with $\kappa = 0$ has a duration of $\pi.$ We neglect
tangential forces in our model and consequently we do not consider
particle rotation. The equations of motion are integrated using a
fourth order Gear predictor-corrector algorithm, with the time step
chosen to be $0.15,$ which still guarantees good energy conservation
in the elastic case.

To prepare the initial configuration, we start from a random
configuration of disks and simulate with periodic boundary conditions
until time $t=20$ using a very low restitution coefficient of $0.1$ to
remove particle overlaps and excess energy in the system efficiently.
Afterwards the setup is continued with $e=1$ until $t=100,$
corresponding to elastic collisions. The average energy per particle
in the system during this stage is quite low and allows only marginal
overlaps.  Thus, the resulting configuration is close to that of a
hard disk system in thermal equilibrium.

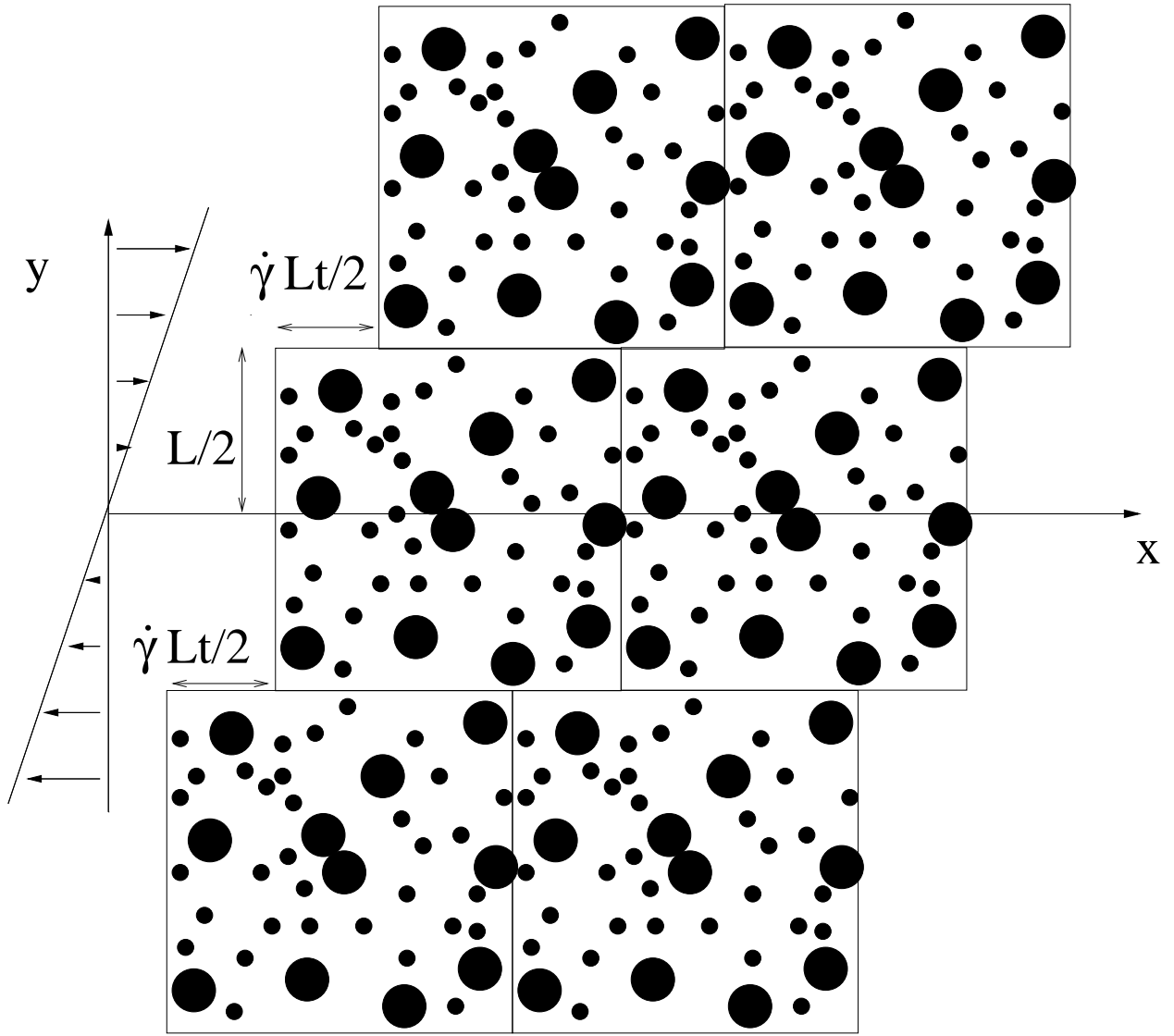
\begin{figure}
%\centerline{\includegraphics[width=.9\textwidth,angle=-90]{sketch.eps}}
%\bigskip
\caption{\label{f:sketch}
  Sketch of the model geometry for shear flow. The simulation cell of
  size $L \times L$ is repeated in $x$-direction (periodic
  boundaries). The cell images in $\pm y$-direction are shifted by an
  amount of $\pm \dot{\gamma}Lt/2$ to reflect the particle
  displacement at the top and bottom of the cell which grows linearly
  in time due to the shear (Lees-Edward boundary condition).
  }
\end{figure}

The liquid motion is approximated by an invariant velocity field in
$x$ direction, 
\begin{equation} \label{e:v-field}
  {\bf u}(y) = y\dot{\gamma}{\bf e}_x, 
\end{equation}
where $y$ is counted from the center of the simulation cell (cf. Fig
\ref{f:sketch}). We have tested the implicit assumption that the
velocity profile is stationary for shear flow, i.e., that it is not
modified by the presence of the particles in the fluid\cite{fn1}.  It
differs only little from the theoretical constant viscosity profile in
the case of Poiseuille flow as shown later in this article.

The liquid exerts an additional drag on each particle $i$ which is 
added to the interaction force (\ref{e:interaction}) between
particles. The force is here assumed to be the Stokes drag force on a sphere,
\begin{equation} \label{e:stokes-drag}
  {\bf f}_{d,i} = - 6\pi r_i \eta \left[{\bf v}_i - {\bf u}(y_i)\right],   
\end{equation}
where ${\bf v}_i$ is the velocity of particle $i$ and $y_i$ its $y$
coordinate.

We then start the simulation by choosing the initial particle
velocities to be equal to the local liquid-background velocity. The
boundary conditions for the particles are Lees-Edwards boundary
conditions (as displayed in Fig. \ref{f:sketch}, cf. also
\cite{b:Allen87}) with shear rate $\dot{\gamma}.$

\section{Results of the shear flow case}

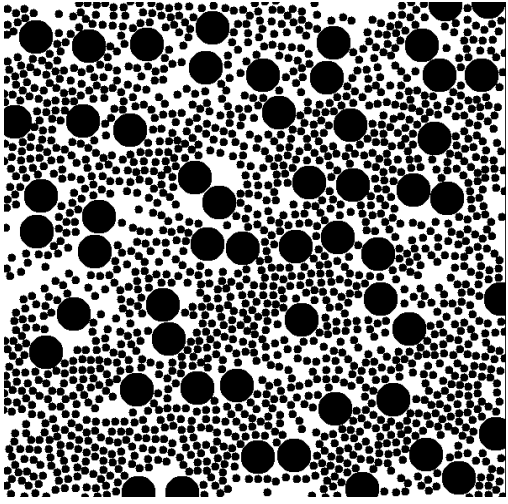
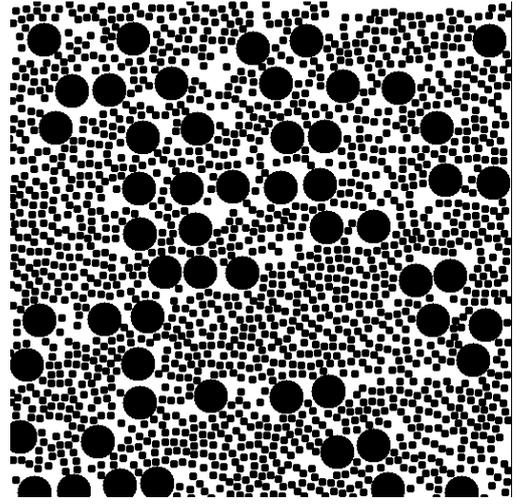
\begin{figure}
%\centerline{
%  \includegraphics[width=.4\textwidth]{ns50.ps}
%  \hfill
%  \includegraphics[width=.4\textwidth]{ns19500.ps}
%  }
%\centerline{\hfill (a)\hfill\hfill (b) \hfill}
%\medskip
%
\caption{\label{f:seq1}
  Simulation snapshots at non-dimensional time $\dot{\gamma} t = 0\ 
  (a)$ and $195\ (b).$ The shear rate in this system is $\dot{\gamma}
  = 0.01,$ the viscosity $\eta = 0.001.$ }
\end{figure}

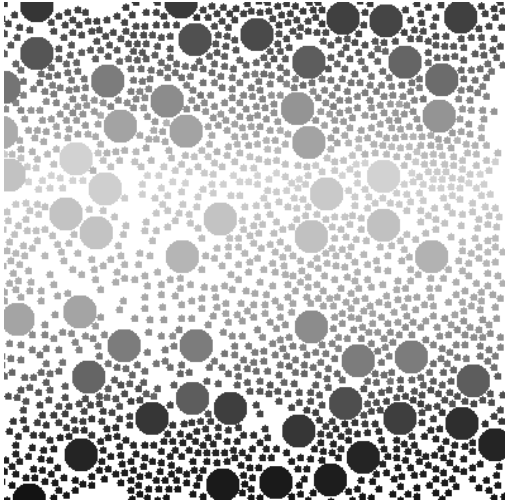
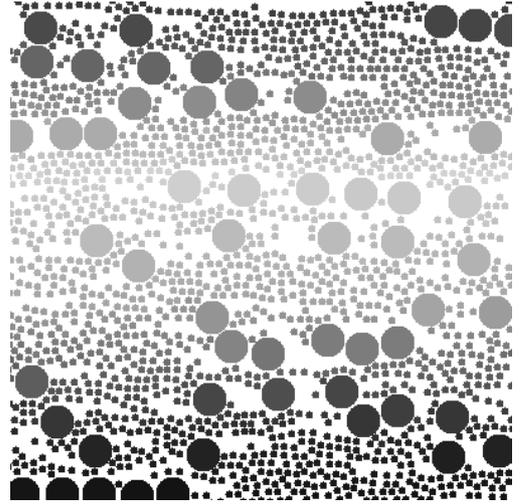
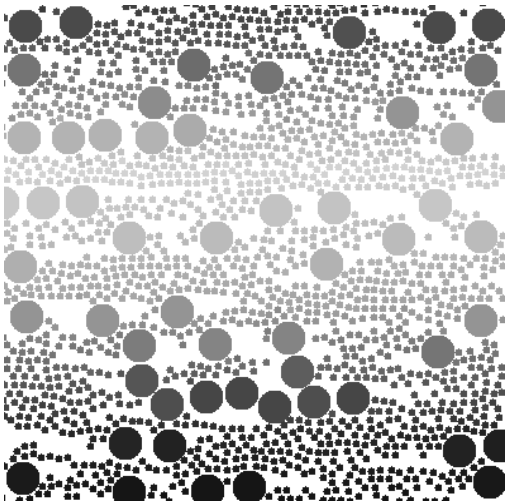
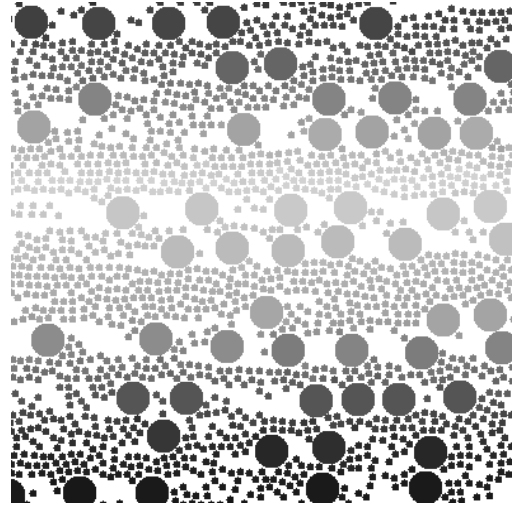
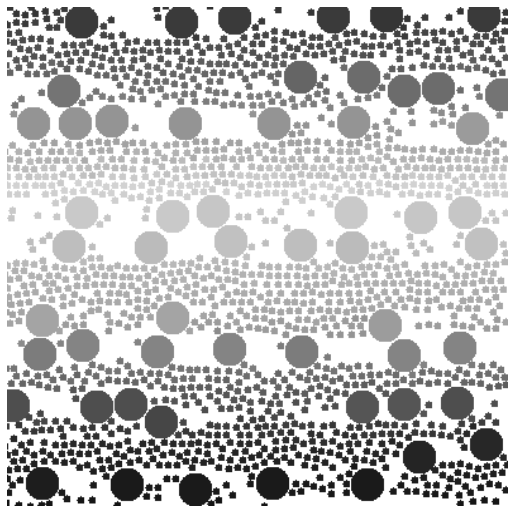
\begin{figure}
%\centerline{
%  \includegraphics[width=.4\textwidth]{s100.ps}
%  \hfill
%  \includegraphics[width=.4\textwidth]{s2000.ps}
%  }
%\centerline{\hfill (a)\hfill\hfill (b)\hfill}
%\medskip
%
%\centerline{
%  \includegraphics[width=.4\textwidth]{s7000.ps}
%  \hfill
%  \includegraphics[width=.4\textwidth]{s11000.ps}
%  }
%\centerline{\hfill (c)\hfill\hfill (d)\hfill}
%\medskip
%
%\centerline{\includegraphics[width=.4\textwidth]{s22000.ps}}
%\centerline{(e)}
%\medskip
%
\caption{
  \label{f:seq2}
  Simulation snapshots at non-dimensional time $\dot{\gamma} t = 0\ 
  (a),\ 20\ (b),\ 70\ (c),\ 110\ (d)$ and $220\ (e).$ The shear rate
  in this system is $\dot{\gamma} = 0.01,$ the viscosity $\eta =
  0.01,$ ten times larger than in the preceding figure. Different
  shades of grey indicate the modulus of the $x$-velocity of the 
  particles.}
\end{figure}

In Figs. \ref{f:seq1} and \ref{f:seq2} we show two typical simulation
sequences whose physical parameters differ only in the value for the
background viscosity $\eta.$ The ratio of the particle radii in our
bidisperse system is $r_b/r_s = 4,$ the ratio of the number density of
big and small particles is $n_b/n_s = 0.05$ and the total area
fraction of particles is rather high, $c=0.6.$ In the long time limit
one sees very different structural reordering of the systems emerging.
In the first sequence (Fig. \ref{f:seq1}) with low viscosity, the
particle arrangement is more or less random, whereas in the second
case (Fig. \ref{f:seq2}) one observes a very clear separation into
alternating zones parallel to the flow direction which contain
alternatingly big and small particles, respectively.

The basis to understand this segregation phenomenon lies in an
analysis of the length scales that are present in the system. Apart
from the system size and the two different particle radii an
additional viscous length exists in the problem. Given that a particle
has a typical particular velocity against the liquid background of
${\bf v}_0$ --- which it acquires in collisions, see below --- the
viscous length $\zeta$ is the typical distance that a particle has to
travel before it again acquires the velocity of the background. The
length $\zeta$ may be estimated in the following way. From the
equations of motion ${\bf f}_{d,i} = m_i \dot{\bf v}$, i.e.,
\begin{equation}
  \label{e:eom-y}
  \dot{v}_y = - \frac{6 \pi r_i \eta}{m_i} v_y, 
\end{equation}
we obtain a simple exponential decay of any excess 
residual $y$ component of the velocity,
\begin{equation}
  \label{e:eom-y-solution}
  v_y(t) = v_0 \exp\left[ - \frac{6 \pi r_i \eta}{m_i} t\right]. 
\end{equation}
with a time constant $\tau = m_i /6 \pi r_i \eta.$ A typical excess
velocity $v_0$ created by a collision of two particles $1$ and $2$ is
$v_0 \approx (\mu/m_i) (r_1+r_2)\dot{\gamma}.$ The product 
\begin{equation} \label{e:zeta1}
\zeta_1 \approx v_0 \tau = \mu (1 + r_2/r_1)\dot{\gamma}/6 \pi \eta 
\end{equation}
estimates the viscous length for particle type 1 after a collision
with a type 2 particle. The value of $\zeta_1$ is largest for
$r_2/r_1=r_b/r_s$ and we call it simply $\zeta$ in the subsequent
text. A detailed discussion is presented in
App.~\ref{a:viscous-length}.

We propose the following physical picture of the segregation process:
The particles move on average with the velocity of the liquid at their
centers. The collisions between particles tend to drive the system to
a state of a random particle distribution\cite{fn2}.  The ratio of the
viscous length to a typical inter-particle distance is a measure for
the efficiency of this process --- the larger the viscous length, the
more effectively will a collision disperse the state of the system.
Conversely however, in a highly viscous environment with $\zeta \ll
1$, collisions will create order.

To see the involved mechanism let us first look for stationary states of a
monodisperse system, which we will define as those states not leading
to particle displacements in $y$-direction of more than half a
particle diameter.  Each single particle defines, due to its finite
extension, a horizontal ``lane'' in the system, i.e., the area that
would be covered as time passes if no other particles were present in
the system. A second particle will undergo collisions with the first
particle if (i) their lanes overlap and (ii) if their $y$-coordinates
differ --- under periodic boundary conditions the particles cannot
escape each other.

For a finite system with a given height one can find a maximum
particle concentration below which all particles can be in disjoint
lanes. But for an infinite system this concentration tends to zero,
since collisions will necessarily occur if the particle arrangements
are random. On the condition that the particle area fraction does not
exceed the maximum packing fraction of disks in a strip one diameter
wide, $c_0 \equiv \pi/4,$ we can imagine a stationary, collisionless
configuration of particles.  Let the particles be arranged one after
another in a horizontal lane, all with the same $y$-coordinate. Then
let these lanes be stacked in $y$-direction. Varying the horizontal
distances between particles and the vertical distances between lanes
any concentration $c \le c_0$ can be achieved.

Such a particle arrangement seems to be highly singular, but it can
under certain conditions be stable against perturbations.  Let the
vertical distance between lanes be small with respect to $r=1$ and
imagine one particle in one of the lanes being slightly displaced in
$+y$-direction. The particle will either undergo a collision with
another particle in the lane above, which will then reduce its
$y$-coordinate again, or it will collide with a particle in its own
lane, increasing the original displacement. If the viscous length
$\zeta$ is large $(\gg 1)$, this collision may suffice to displace
both collision partners to another lane (if at the same time also the
mean free path is large enough, see App.~\ref{a:mean-free-path}).  In
contrast, if $\zeta$ is small $(\ll 1)$, then the relative motion of
the two particles is more akin to a sliding on top of each other,
displacing each particle by $\approx 1/2$ in $+y$ and $-y$-direction
respectively. In a sufficiently dense system subsequent collisions
with particles can the neighboring lanes can then restore the original
vertical positions of the particles.

In the polydisperse case we might now expect that mixed lanes,
containing both big and small particles, are stable. However,
arguments can be made against this supposition. If $\zeta$ is small in
comparison to the big particle radius then small particles in lanes
containing big particles will be ejected from the big particle lane by
collisions.  The reason is that the big massive collision partner is
not significantly displaced by one collision, whereas the small
partner is. If then the neighboring lane contains small particles, it
is not hard to absorb the additional expelled particle.  If, however,
the neighboring lane contains big particles, alternating collision
series of the small particle with big particles below and above will
establish an additional small particle lane in between the big
particle lanes.
  
%Due to the larger cross section of the big particles, collisions between big
%and small particles are more likely than collisions between two small
%particles.  If the viscous length is smaller than the radius of the
%big particle, then the big particle acts just as some sort of snow
%plough and shuffles small particles out of its way.  The fast small
%particles colliding with its upper half will obtain velocity
%components in $+y$ direction, and the slower small particles colliding
%with its bottom half will get gain velocity in $-y$ direction. After
%possibly some collisions the small particles will come to rest with
%respect to the liquid at a distance of at most $\zeta$ below or above
%the big particle.  

Starting from a random configuration at small $\zeta,$ local groups of
big particles will tend to align and expel small particles ---
neighboring groups may then join due to the stresses generated by the
collisions with the small particles accumulating outside these groups.
Finally, stable lanes form segregates into a of particles in disjoint
lanes.  A typical simulation sequence of this ordering process is
displayed in Fig.~\ref{f:seq2}.

To arrive at a quantitative description of the segregation process, we
define an ``order parameter'' in the following way. Since we have
observed a strong stratification of the flow into horizontal layers,
we define for each particle species the area fraction in a horizontal
strip of width of the average particle radius. For a completely
segregated system, we expect the area fraction of small particles
$a_s$ to be large whenever the fraction of big particles $a_b$ is
small.  Consequently, the quantity
\begin{equation}
  \label{e:delta}
\delta \equiv \langle \left[a_b-a_s-\langle a_b-a_s\rangle\right]^2
\rangle^{1/2}  
\end{equation}
is small for a random mixture of the two species and assumes a large
value when the system is stratified. %\footnote{
%
%The definition (\protect\ref{e:delta}) will yield slightly smaller
%values of $\delta$ when the systems become wider. In this case it 
%may be necessary to further partition the system vertically. 
%}
%
The brackets denote the average over all examined horizontal slices. 

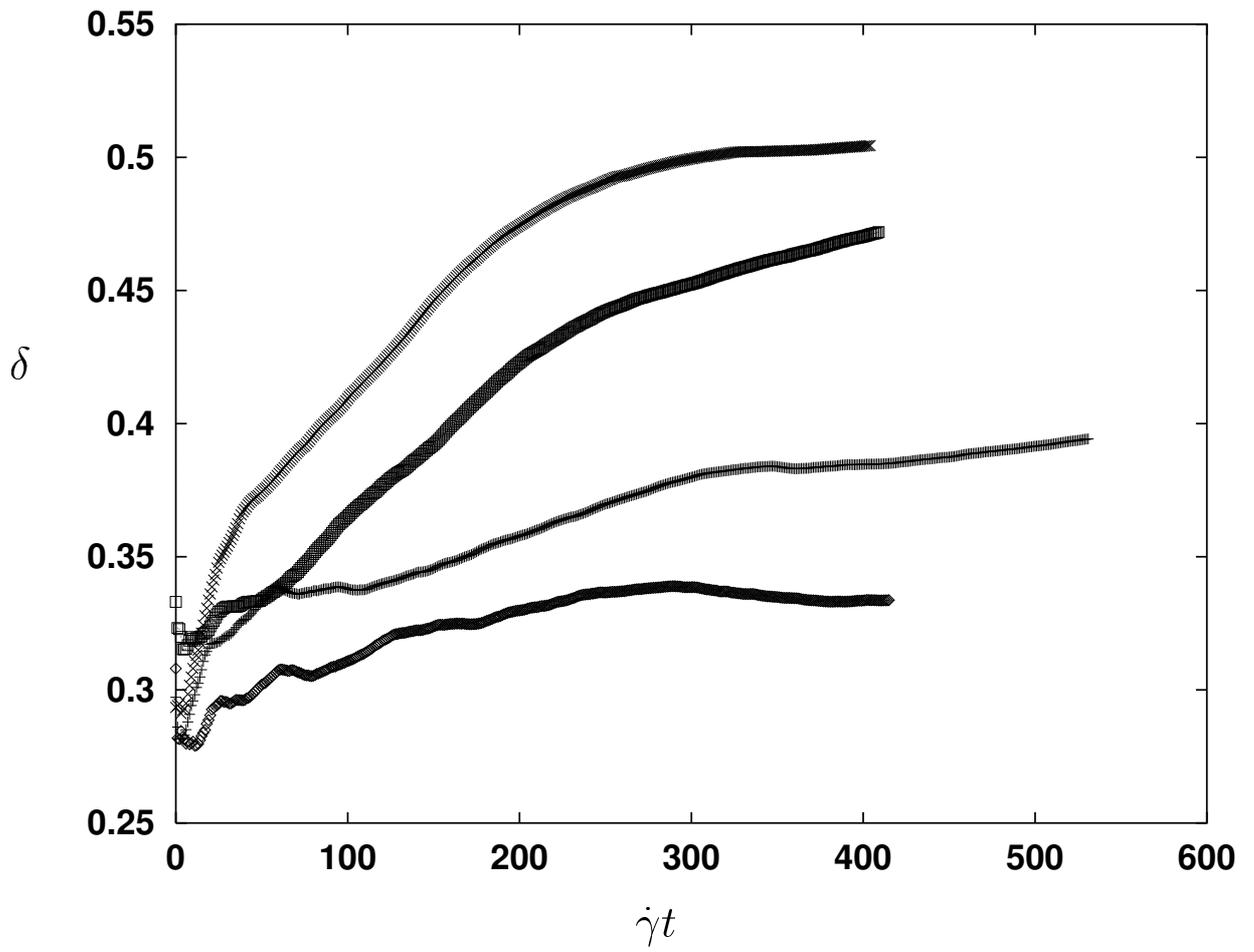
\begin{figure}
%\unitlength=\textwidth
%\begin{picture}(1,0.8)
%\put(0,0.8){\includegraphics[width=.7\textwidth,angle=-90]{delta_t.ps}}
%\put(0,0.5){\Large $\delta$}
%\put(0.5,0.05){\Large $\dot{\gamma}t$}
%\end{picture}
%\medskip
\caption{\label{f:delta_t}
  Time dependence of the segregation parameter $\delta$ for
  simulations with different viscosity $\eta = 0.001,\ 0.004,\ 0.01,\ 
  0.03$ (bottom to top) corresponding to $\zeta \approx 2.5,\ 
  0.63,\ 0.25,\ 0.08$ [according to Eq. (\protect\ref{e:zeta-final})]
  (bottom curve to top curve) vs. dimensionless time $\dot{\gamma}t$
  on the abscissa.}
\end{figure}

In Fig.~\ref{f:delta_t} we show the time dependence of $\delta$ for
different values of $\zeta$ for constant overall area fraction $c=0.6$
and constant shear rate $\dot{\gamma} = 0.01.$ We clearly see an
initially fast size segregation process which becomes slower and
slower and finally saturates to a value $\delta_\infty$ that depends
on $\zeta.$ Due to the rather low value of $\zeta$ and consequently a
strong non-ergodicity of our systems, the sample-to-sample
fluctuations are large and values of $0.2\delta_\infty$ are typical in
samples of size $100 \times 100.$

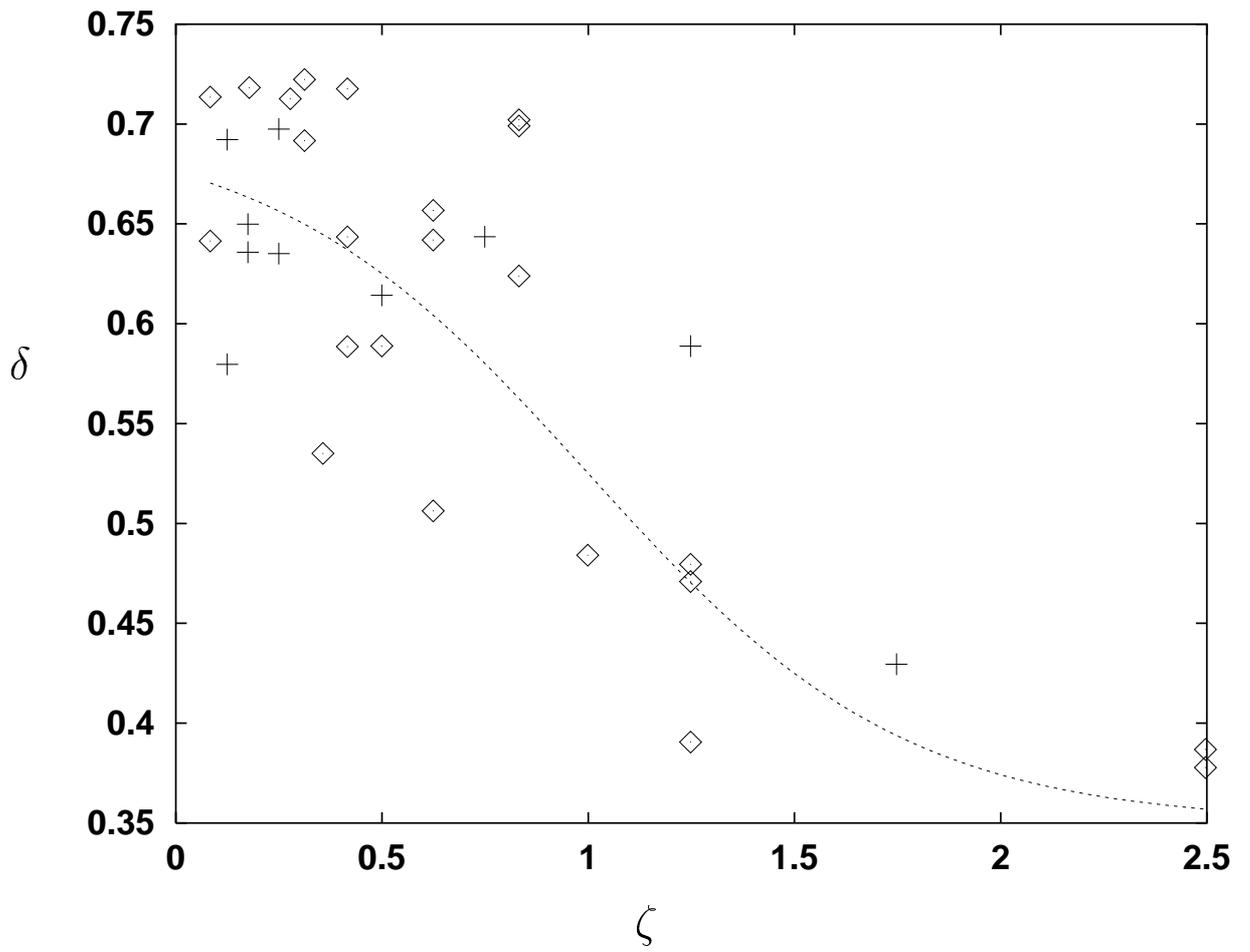
\begin{figure}
%\unitlength=\textwidth
%\begin{picture}(1,0.8)
%\put(0,0.8){\includegraphics[width=.7\textwidth,angle=-90]{collapse.ps}}
%\put(0,0.5){\Large $\delta$}
%\put(0.5,0.05){\Large $\zeta$}
%\end{picture}
\caption{\label{f:collapse}
  Final values of the segregation parameter $\delta$ plotted vs.
  viscous length $\zeta$ [according to Eq.
  (\protect\ref{e:zeta-final})] on the abscissa for several values of
  viscosity and shear rate in the system. Crosses denote computations
  with constant viscosity $\eta=0.01$, diamonds indicate constant
  shear rate $\dot{\gamma}=0.01.$}
\end{figure}

The decreasing dependence of $\delta_\infty$ on $\zeta$, which
demonstrates the mixing or destabilizing effects of large $\zeta,$ is
shown in Fig. \ref{f:collapse}. The figure shows data obtained for
different fluid viscosities and shear rates $\dot{\gamma},$ but
constant overall area fraction. The scatter is rather large due to the
abovementioned sample-to-sample fluctuations.  At large $\zeta$, the
segregation does not increase significantly over the initial value. In
fact, if $\zeta$ is larger than the mean free path between particles,
then the spatial distribution of particles does not differ much from
that of the corresponding, inelastic, sheared hard-core gas
\cite{Goldhirsch95}. However, at small $\zeta$ the friction with the
liquid is very large and causes ordering in bands oriented parallel to
the flow containing alternatingly big and small particles.  For a
proposal to scale not only with respect to $\dot{\gamma}$ and $\eta$
but also incorporating the concentration $c,$ see
App.~\ref{a:mean-free-path}.

\section{Poiseuille flow}

To see the influence on the ordering effect of a different flow
profile, we have exchanged the simple linear shear profile of
Eq.~(\ref{e:v-field}) by a quadratic Poiseuille profile,
\begin{equation} \label{e:v-poiseuille}
  {\bf u}(y) = 4L (\frac{y}{L}+\frac{1}{2})(\frac{y}{L}-\frac{1}{2}) v_{\rm max} {\bf e}_x. 
\end{equation}
We keep periodic boundary conditions in $x$-direction while
introducing reflecting walls at the bottom and the top of the cell,
where the flow velocity is zero in accord with no-slip boundary
conditions.  The shear rate now depends on the $y$-coordinate in the
flow,
\begin{equation} \label{e:v-poiseuille-shear}
  \dot{\gamma}_y = 8 \frac{y}{L} v_{\rm max}. 
\end{equation}
Its modulus $|(\partial/\partial y) v_x|$ is largest at the walls and
zero in the center of the flow.

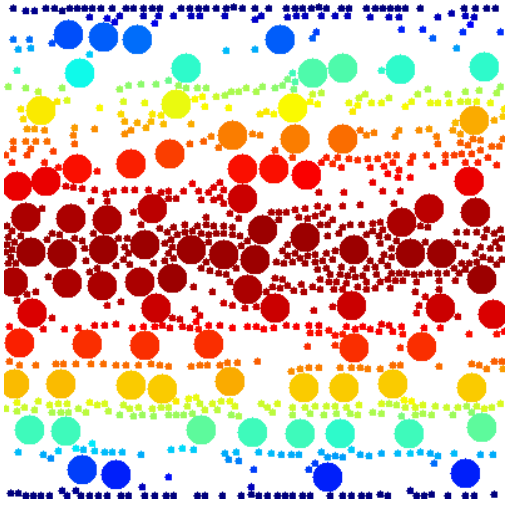
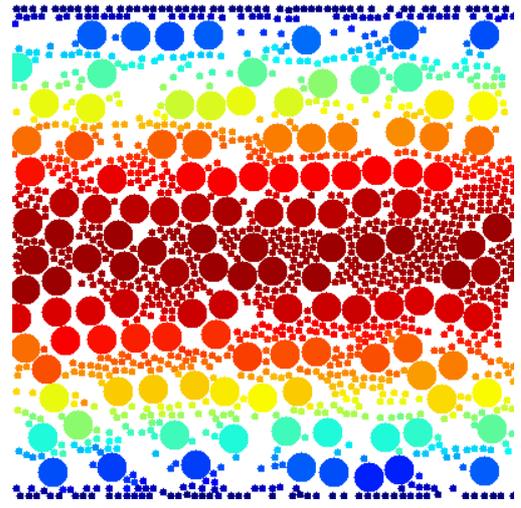
\begin{figure}
%\centerline{
%  \includegraphics[width=0.4\textwidth]{pois-0.4.ps}
%  \hfill
%  \includegraphics[width=0.4\textwidth]{pois-0.6.ps}
%  }
%\centerline{\hfill (a)\hfill\hfill (b) \hfill}
%\bigskip
\caption{\label{f:pois_c}
  Stationary particle configuration with imposed parabolic Poiseuille 
  flow profile. The system size is $100 \times 100,$ the particle
  volume fraction $c=0.4\ (a),$ and $0.6\ (b).$ }
\end{figure}

We have studied systems of concentration $c=0.6$ and $c=0.4$ with
viscosity $\eta = 0.01$ and shear rates of $\dot{\gamma}_{L/4} = 0.01$
at $y= L/4$ so that a typical zone in the flow corresponds to the
crossover regime where $\zeta \approx 1.$ After a long time
$t\dot{\gamma}_{L/4} \approx 100$ we find that the particles
concentrate in the central region of the cell, where the shear rate is
smallest, cf. Fig.  \ref{f:pois_c}. Close to the wall, and although
the shear is largest there, we still find stratification --- small
particles gather at the wall.

%The increase of the number of small particles in zones of low shear is
%due to the presence of a range of shear rates in the system.  Note
%that in fact the shear rate sets a second physical time scale in the
%flow, namely the time that the system requires for a differential
%horizontal displacement of unit length per unit length in vertical
%direction. For a given particle concentration, this time scale
%determines the collision frequency between particles. Now we have seen
%in the shear flow case that the large particles tend to displace the
%small ones. Thus, the displacement rate of the small particles is an
%increasing function of the shear rate --- given the viscosity is high
%enough to place $\zeta$ in the segregation regime. Thus the high shear
%regions close to the walls will ``expel'' their small-particles
%content very fast and these will be collected in the low shear zones
%of the flow. Although $\zeta \to 0$ where the flow speed is maximal
%and thus segregation should occur, the segregation effect is
%counterbalanced by the flow of particles originating in the high shear
%regions.

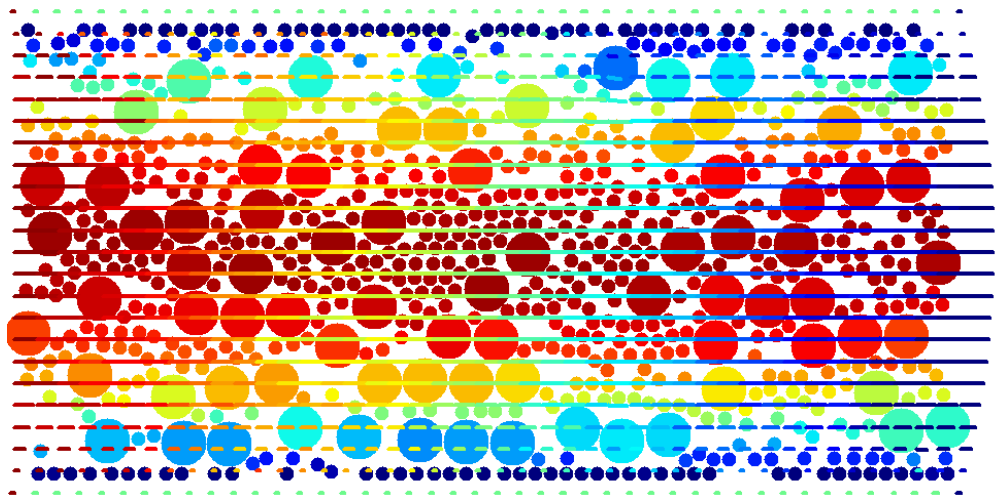
\begin{figure}
%\centerline{\includegraphics[width=0.8\textwidth]{pois-0.6-ns.ps}}
%\bigskip
\caption{\label{f:poiseuille}
  Final particle configuration in Poiseuille flow with velocity
  profile obtained by solution of the full Navier-Stokes equation with
  a point force term modeling the momentum exchange between fluid and
  particle phase.  There is no difference in the physical parameters
  of the two systems. The system size is $80 \times 40.$ }
\end{figure}

In a real experimental system, the liquid is strongly coupled to the
particles. The momentum exchange between the two phases alters
significantly the flow pattern, both on the scale of the particle
radius, but at least in the viscous regime also on the scale of the
container. We have tested our assumption of a laminar profile on the
scale of the container by comparison to the results of a newly
developed algorithm whose details are described in
Refs.~\cite{Kalthoff95,Schwarzer95}. There, we couple the particles to
the liquid by Stokes-type drag forces proportional to the particle
radius and the local velocity difference to the flow field ---
corresponding to Eq. (\ref{e:stokes-drag}). The drag is introduced as
a point-force term in the Navier-Stokes equations for the liquid
surrounding the particles and accounts for the momentum transfer from
the particles to the liquid. The Navier-Stokes equation is solved by a
finite difference scheme, where the grid has a resolution on the order
of the particle size. This technique is appropriate to observe the
long range effects of the hydrodynamic interaction between the
particles. The result of this simulation is shown in Fig.
\ref{f:poiseuille}, for the same choice of parameters as for the fixed
velocity profile.

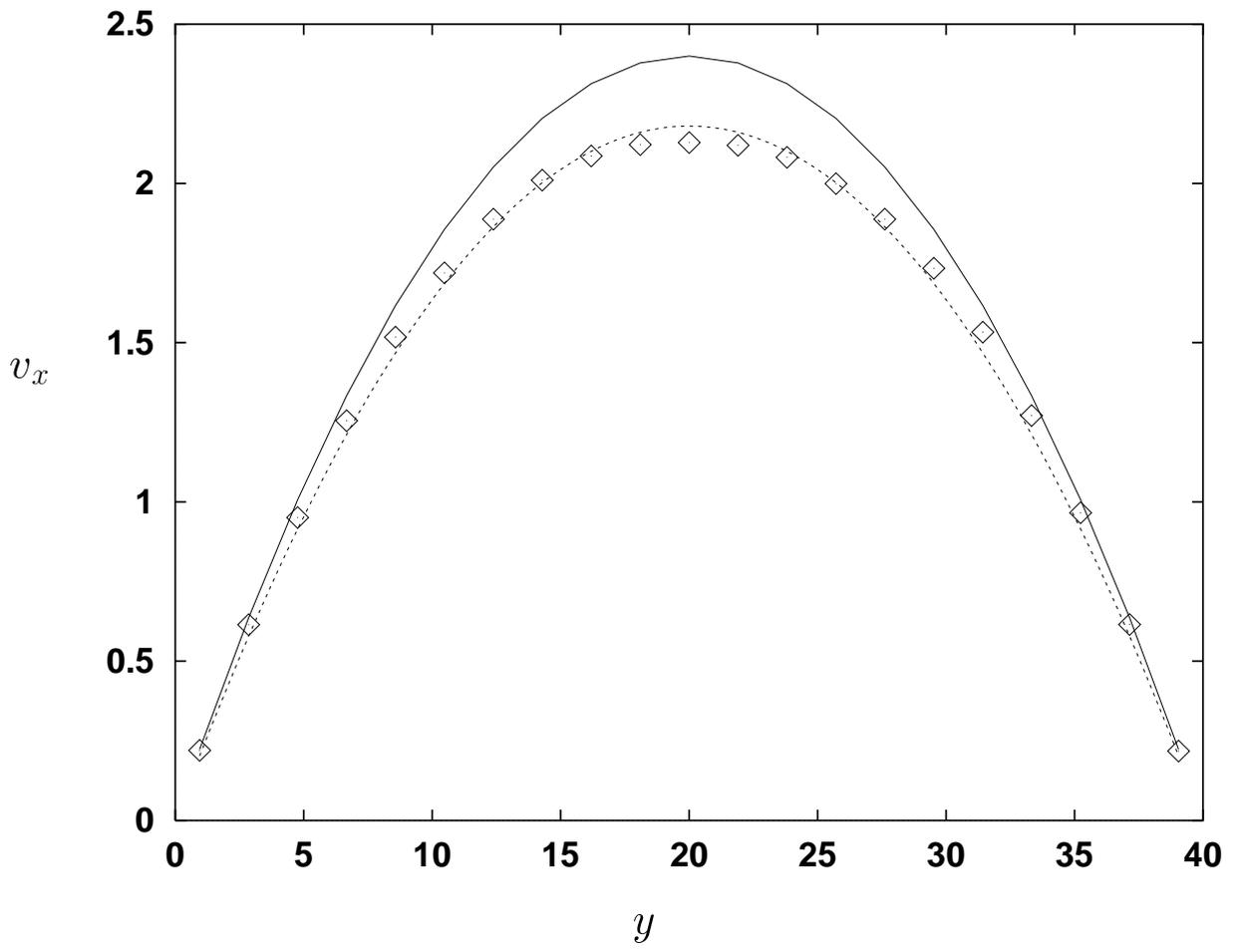
\begin{figure}
%\unitlength=\textwidth
%\begin{picture}(1,0.8)
%\put(0,0.8){\includegraphics[width=0.7\textwidth,angle=-90]{profile.ps}}
%\put(0,0.5){\Large $v_x$}
%\put(0.5,0.05){\Large $y$}
%\end{picture}
%\medskip
\caption{\label{f:profile}
  The $x$-component of the velocity profile along a cut in
  $y$-direction in the Navier-Stokes Poiseuille flow simulation at
  $\mbox{Re} = 100.$ The solid line is the prediction for laminar flow
  without particles, the dashed line is a parabolic fit to the data
  points.}
\end{figure}

Since there is no drastic visible difference to the fixed profile
calculation shown in Fig.~\ref{f:pois_c}(b) with the same physical
parameters, we have {\it a posteriori\/} justified the choice a fixed
liquid profile. To be more quantitative, we have checked the velocity
profile of the combined particle-liquid system against the theoretical
expectation for laminar liquid-only flow (see Fig.~\ref{f:profile})
with constant viscosity. For moderate Reynolds numbers $(\approx 100)$
we find that there are only very small deviations from a parabolic
flow profile.  However, the particle load increases the effective
viscosity of the mixture and thus the overall flow velocity is reduced
in comparison to the case of the pure liquid. In Fig.~\ref{f:profile}
we recognize a slight flattening of the parabolic profile near the
center of the flow. Such a behavior is characteristic for shear
thinning liquids and is here caused by the locally increased viscosity
in the flow center, corresponding to the increased vertical momentum
transport in the denser central region.  At these Reynolds numbers, we
measure only small velocity fluctuations of the liquid along the
horizontal direction, i.e., the profile is stationary.

\section{Conclusion}

We have studied sheared bidisperse granular systems under conditions
of simple shear and Poiseuille flow. We find that the presence of the
liquid can induce particle size segregation. The criterion to decide
whether this segregation mechanism will occur is that the ratio of
shear rate to viscosity must be small. This statement is equivalent to
saying that the viscous length in the system should be small when
compared to a typical linear scale in the problem as, e.g., the
particle radius. The particles arrange themselves in bands moving with
the flow that contain alternatingly big and small particles.

Also in Poiseuille flow we observe the formation of bands of particles
of different size.  Here, in addition, the variety of time scales in
the flow leads to an aggregation of particles in the zones of low
shear rate. These results have been verified against simulations using
a full Navier-Stokes simulation for the liquid.

The proposed segregation mechanism relies on the presence of a liquid
phase. It is thus very different from known mechanisms in dry granular
media, where gravity induced avalanches occur and separate particles
species whose static angles of repose differ.

It should be very interesting to study the behavior of the system with
more than two particle species or a whole continuum of species or the
dependence on the particle radius ratio. Moreover, three dimensional
calculations are desirable and quantitative comparisons to experiments
are necessary.

\section{Acknowledgments}

We would like to thank in particular Stephane Roux for valuable
suggestions. St.~S.\ thanks Wolfgang Kalthoff and Stefan Luding 
for discussions.
%This
%work has been motivated by studies financed by a grant from the
%Groupement de recherche XXX. 
We thank the H\"ochleistungsrechenzentrum
of the Forschungszentrum J\"ulich for computer time on their Paragon
parallel computer. St.~S.\ has been supported by a scholarship from the
scientific council of NATO, granted by the Deutscher Akademischer
Austauschdienst, Bonn.

\appendix

\section{Viscous length}
\label{a:viscous-length}

We obtain an estimate (when no further collisions occur) for $\zeta$
by integration of the equations of motion ${\bf f}_{d,i} = m_i
\dot{\bf v}.$ Taking the drag force from Eq.~(\ref{e:stokes-drag}), we
have obtained for the $y$ component of the equation of motion
(\ref{e:eom-y}), which we print here once more:
\begin{equation}
  \label{e:a-eom-y}
  \dot{v}_y = - \frac{6 \pi r_i \eta}{m_i} v_y. 
\end{equation}
This relation leads to a simple exponential relaxation of the initial excess
velocity $v_y(0),$
\begin{equation}
  \label{e:a-eom-y-solution}
  v_y(t) = v_y(0) \exp\left[ - \frac{6 \pi r_i \eta}{m_i} t\right]. 
\end{equation}
In analogous fashion, we obtain the expression for the $x$ component,
\begin{equation}
  \label{e:eom-x}
  \dot{v}_x = - \frac{6 \pi r_i \eta}{m_i} 
                \left[ v_x - \dot{\gamma}\int_0^t dt'\ v_y(t') \right],
\end{equation}
which we can integrate by standard methods to find $v_x(t)$. We define
the viscous length as the norm of the vector valued integral  
\begin{equation}
  \label{e:xmu-def}
  \zeta = \left\| \int_0^\infty dt \ 
    \left[{\bf v}(t) - {\bf u}(y(t))\right] \right\|
\end{equation}
and use the solutions of (\ref{e:a-eom-y}) and (\ref{e:eom-x}) to
obtain, 
\begin{equation}
  \label{e:xmu}
  \zeta = \frac{m_i}{6 \pi r_i \eta} \left\| \left( 
        \begin{array}{c} v_x(0) -
          \frac{\dot{\gamma} m_i v_y(0)}{6 \pi r_i \eta} \\
          v_y(0) 
        \end{array} \right) \right\|. 
\end{equation}
We then find typical values for ${\bf v}_0$ by a consideration of a
two particle collision --- say between particles with label 1 and 2
--- assuming that the initial velocities equal ${\bf v}_i = {\bf
  u}(y_i)$ according to their different $y$ positions in the flow (for
a more complete discussion of inelastic two particle collisions, see,
e.g. \cite{Walton92}). The velocities after the collision, in the
reference frame comoving with the liquid at the initial position of 
particle 1, are
\begin{equation}
  \label{e:v0}
  {\bf v}_1(0) = - \frac{\mu}{m_i}b\dot{\gamma} \left( 
    \begin{array}{c} \sin \alpha \cos \alpha (1+e) \\
                     \sin^2\alpha - e\cos^2\alpha + 1 
    \end{array}  \right).
\end{equation}
Here, $b$ denotes the impact parameter and $\sin \alpha \equiv b/
(r_1+r_2)$, cf. Fig. \ref{f:impact}. Thus, apart from order-one
geometrical factors and some $e$ dependence, the velocity of the
scattered particle is\cite{fn3} $\approx (\mu / m_i) ( r_1 + r_2 )
\dot{\gamma}$. Therefore, $\zeta$ becomes largest for the small
particles after a collision involving a big and a small particle:
\begin{equation}
  \label{e:zeta-final}
  \zeta \approx \frac{\mu (1 + r_b/r_s)\dot{\gamma}}{6 \pi \eta}.  
\end{equation}

\section{An estimate for the mean free path in the monodisperse system}
\label{a:mean-free-path}

The ratio of the viscous length to the mean free path of the particles
in a non-viscous environment is probably an important dynamical
characteristics of the system. If the mean free free path is much
shorter than the viscous length, the additional background viscosity
will not have a big effect. If on the other hand the mean free path is
much longer than the viscous length, then the behavior of the system
is viscosity dominated.  Here, we would like to give an estimate of
the mean free path in a monodisperse system for particles moving in
the vertical direction. To allow for a simple calculation in the
stationary situation, we resort to a simple hypotheses for the system's
configuration at large times: due to the initial disorder the systems
arranges itself such that the number of horizontal particle lanes is
maximum.

We note that under these circumstances the average horizontal distance
$d$ between two particles is set by the packing fraction. If the
number of lanes is maximum, their width must be the smallest possible,
namely $2r.$ One particle covers an area of $\pi r^2$ within the
available area $2rd$. Consequently, the overall area fraction is
\begin{equation}
c=\frac{\pi r^2}{2rd} = \frac{1}{2}\pi \frac{r}{d}. 
\end{equation}

As in the previous section, we now assume that we perturb the
trajectory of one particle by giving it a vertical velocity of order
$r\dot{\gamma},$ which is of the same order as the velocity difference
between two lanes $2r\dot{\gamma}.$ If $d = 2r,$ the system will not
allow particles to penetrate into the neighboring lane. This situation
is the densest packing compatible with a stationary state of the
system, $c_0 = \pi/4.$ As long as $d < 4r,$ or equivalently, $c > c_1
= \pi/8,$ a particle will only occasionally be able to pass a lane.
Considerations of the particle geometry apart, the probability for a
hit should be proportional to the time spend in the lane by the
scattered particle divided by the average time between the pass of two
successive particles in the neighbor lane. The mean free path $\ell$
is given by the condition that this ratio be about 1, i.e.,
\begin{equation}
1 \approx \frac{\ell/\dot{\gamma}r}{(d-2r)/2\dot{\gamma}r,}
\end{equation}
or, 
\begin{equation} \label{e:a2-high-c}
\ell/r \approx (d-2r)/2r = \frac{1}{4}\pi c^{-1} - 1 = 
       \frac{c_0}{c} - 1. 
\end{equation}

For even lower concentration, the particle has a good chance to pass
one or even several lanes, each with probability $1-p_{\rm hit}\approx
1 - [2r / \dot{\gamma}r]/[(d-2r)/2\dot{\gamma}r] = 1- (2r)/(d-2r).$
The probability to survive a distance $x/r$ without hits is hence
distributed exponentially,
\begin{equation}
p(x/r) \sim (1-p_{\rm hit})^{x/2r}, 
\end{equation}
which yields --- by normalization and determination of the expectation
value ---
\begin{equation}
\ell / r =  - 1/\ln(1-p_{\rm hit}).
\end{equation}  
For small concentrations (and thus also small hitting probabilities)
this equation may be expanded to yield the same form as
(\ref{e:a2-high-c})
\begin{equation} \label{e:a2-low-c}
\ell / r \sim  1/p_{\rm hit} \sim \frac{c_0}{c} - 1. 
\end{equation}  

It is interesting to note that $\ell$ does not depend on the
shear rate but only on geometrical properties of the system.  This
observation is in favor of our suggestion that the ratio of viscous
length to particle radius $\zeta/r$, as proposed in the main text and
here in dimensional form, collapses the simulation data for $\delta$
at a fixed given $c$.  If $\dot{\gamma} \ll 1$, we presume that
$\zeta/\ell$ may be a good scaling variable, even at different area
fractions.  This may be an interesting question to investigate.

%\bibliography{strings,physics,physics-na,books,own} 

\begin{thebibliography}{10}
\bibitem{fn1} Of course the flow will differ on the scale of a
  single particle --- we mean here that the flow remains unaltered on
  the scale of several particle sizes.
\bibitem{fn2} The particle distribution will not become truely
  random, because dissipative systems form particle
  ``clusters''\cite{Goldhirsch93} and in the elastic case the system
  heats up due to the applied shear and will have a particle
  distribution close to that of the hard-core gas.
\bibitem{fn3} We would like to point out that even in more
  elaborated models, which may include a more realistic rendering of
  the hydrodynamic interaction between the particles, the relevant
  velocity scale is of order $b\dot{\gamma}$ so that the following
  arguments remain valid.
\end{thebibliography}

\newpage

%\newpage

%\newpage 

%\newpage 

%\newpage

%\bigskip
%\bigskip

\end{document}